\documentstyle[12pt,epsfig,amssymb,lscape]{article}
\begin{document}
\title{The string spectrum from large Wilson loops}
\author{Pushan Majumdar \thanks{email: pushan@mppmu.mpg.de} \\
Max-Planck-Institut f\"{u}r 
Physik, \\ F\"{o}hringer Ring 6, D-80805 M\"{u}nchen, Germany.} 
\date{22 November 2002 \\ MPI-PhT/2002-70}
\maketitle
\begin{abstract}
We look at energies of the low lying states of the hadronic string in three 
dimensional SU(2) lattice gauge theory by forming correlation matrices among 
different sources. We are able to go to previously inaccessible time separations. 
This is made possible by using a new algorithm proposed by L\"{u}scher and Weisz 
which lets us measure the exponentially small values of large Wilson loops 
with sufficient accuracy.   
\end{abstract}
\section{Introduction}

The mechanism of quark confinement in Quantum Chromodynamics remains 
unsolved to this date. One of the most appealing pictures of confinement is 
that in the QCD vacuum, flux tubes are formed between quarks and anti-quarks 
giving rise to a linearly rising potential.

There have been several attempts \cite{strings} to write down effective theories for 
these flux tubes also known as hadronic strings. For a recent review see \cite{strrev}. 
Typically in all these theories, the potential is 
represented as a series in $r$ (the quark anti-quark separation) with $\sigma r$
as the leading term at large $r$. The coefficient $\sigma $ of this linear term is 
called the
string tension. The sub-leading terms go as inverse powers of $r$. One of 
the most striking features of these theories has been the prediction of a universal 
coefficient of the $1/r$ term \cite{uni}. This coefficient,
 commonly denoted as $c$, has the value
$-\frac{\pi}{24}(d-2)$ where $d$ is the number of space-time dimensions.
Several studies have looked at features of the string \cite{matel} \cite{strfeat} and, 
in particular, the ground state and the coefficient $c$ \cite{LW1} are known quite well.   
However controversy still exists regarding the excited states and 
the predicted energy differences of integral multiples of $\pi/r$. A recent study 
by Caselle {{\em et al.} 
\cite{Ising} looks at the string picture in Ising gauge theories using 
powerful
 numerical techniques which uses the dual symmetry of the model. However for 
non-Abelian gauge theories, such precise numerical techniques are lacking. To 
circumvent this problem, studies by Juge {\em et al.} 
\cite{asymm} \cite{asymm1} \footnote{ We thank the referee for pointing out this 
reference.} on Yang-Mills theories have relied on asymmetric lattices.

A recent algorithm suggested by L\"{u}scher and Weisz \cite{multi} go a long way to
remedy this problem as it allows us to measure small expectation values
with good accuracy even for theories with continuous gauge symmetries. For a 
discussion of the algorithm and optimization 
issues relevant for Wilson loops we refer the reader to \cite{optim}. 
Recently this algorithm has been used to look at string breaking \cite{F1},
static 3-quark potential \cite{F2} and correlation between pairs of Wilson 
loops \cite{Meyer}. Here we use this algorithm to measure the low lying
states of the hadronic string in pure Yang-Mills lattice gauge theory.

We would like to point out that this study is only exploratory in nature. 
Therefore we have concentrated more on measurements at larger time separations 
using the L\"{u}scher - Weisz algorithm rather than using sophisticated wave 
functions. A complete study will probably require both. Due to computational 
and time constraints we were unable to undertake such a study at the moment,
but hope to do so in the future.

Section II is devoted to setting the notation and a discussion of lattice 
preliminaries.
In section III we look at the classification of the string states. Section
IV is devoted to some of the simulation details. In particular we set the
scale on the lattice and discuss the smearing scheme in this section. In section
V we present our results on the spectrum and discuss different procedures
to extract the energy of the states. Section VI deals looks at the energy 
differences at different $\beta$ values, and finally in section VII we draw our 
conclusions. In the appendix we present a discussion of our error analysis.

All our figures, unless explicitly stated, are in lattice units.

\section{Preliminaries}

We will work in 2+1 dimensions with SU(2) lattice gauge theory and the Wilson
action. We also impose the usual periodic boundary conditions. The fundamental 
degrees of freedom of this theory are the SU(2) matrices associated
to the links of the lattice which we denote by $U$. The action is defined on 
the smallest closed paths on the lattice called plaquettes.
Our partition function is given by
\begin{equation}
Z=\int_{U \in SU(2)}{\cal D}U \; e^{\frac{\beta}{2} \sum_p tr (U_p)}
\end{equation}
where the sum is over all plaquettes $p$ and $U_p$ is the directed product of 
$U$'s around the plaquette.
To make a connection to the continuum theory we have to identify $\beta$
with $4/g^2$ and take both $g$ and the lattice spacing to zero.
The choice of Wilson action is also important for us as this action allows
the construction of a positive transfer matrix \cite{Luscher}. Both the 
algorithm that we use, and the physical interpretation of what we do, rely on 
the existence of such a
transfer matrix. 

The transfer matrix of a model provides the relation between 
the functional integral and the Hamiltonian formalism. Let us call links in the
space direction as $U_s$ and in the time direction as $U_t$. 
In the temporal gauge where
all the time like links $U_t(t+1,t)$ are set to ${\bf 1}$,
it is easy to interpret the transfer matrix as connecting states in one
time slice to the next. The states on which this
 transfer matrix acts are square integrable wave functions $\Psi[U_s]$.
The space of these wave functions form a Hilbert space.

The partition function $Z$ in the presence of external test charges can be 
written in terms of the transfer 
matrix as $$Z_{{\bar q}q}=tr({\mathbb P}_{{\bar q}q}{\cal T}^n)$$ where
${\mathbb P}_{{\bar q}q}$ is the projection onto the relevant sub-space
of the Hilbert space. For a more complete discussion of the transfer
matrix formalism see \cite{Seiler}.

The main observable that we are concerned with is the Wilson loop. In 
the continuum, the Wilson loop is given by
\begin{equation}
W(C)=tr\;{\bf P}\left ( \exp(\int_C A_{\mu}dx_{\mu})\right )
\end{equation}
where $C$ is any closed curve and {\bf P} denotes path ordering.
On the lattice the Wilson loop is defined by 
\begin{equation}
W(C)=\frac{1}{m}\;tr\{\prod_{l\in C}\;U_l\}
\end{equation}
where again $C$ is a closed curve and $m$ is the dimension of the matrix
$U$.

Wilson loops of extent $(R,T)$ can be interpreted in the transfer matrix formalism, 
as the correlation of sources between points separated by a 
distance $R$, propagating for time $T$.
Thus it is possible to look at 
correlations between string states by choosing appropriate
$\Psi[U_s]$ as sources for the Wilson loops in Yang-Mills theory.

The other observable that we look at is the Polyakov loop. It is defined by 
\begin{equation}
P({\vec x},T)=tr\;{\bf T}\left ( \exp(\int_0^T A_0({\vec x},t)\;dt)\right )
\end{equation}
where $T$ is the extent of the lattice in the time direction and {\bf T} 
denotes time ordering.
On the lattice the Polyakov loop is defined by
\begin{equation}
P({\vec x},T)=tr\{\prod_{l=0}^T\;[U_t({\vec x})]_l\}.
\end{equation}
The Polyakov measures the excess free energy of the vacuum induced 
by a static test quark.

\section{String states}

The states of the hadronic string that we want to look at carry
different quantum numbers. Therefore it is important to classify them.
 Here we briefly outline the classification scheme.
 
The string states that we are interested in, are configurations 
at fixed time with both ends of the string fixed so 
that only transverse degrees of freedom are left.
In analogy to \cite{dual}, an effective Hamiltonian for the transverse 
degrees of freedom for a string of length $r$ can be written as 
\begin{equation}
{\cal H}=\frac{\pi}{2r^2}\int_0^r\,d\kappa \left ( \frac{{\cal 
P}_i^2(\kappa)}{\sigma}
+\sigma x_i^{\prime 2}(\kappa) \right )
\end{equation}
where $x^{\prime}$ is derivative of $x$ with respect to $\kappa$
and ${\cal P}$ are the canonical momenta. $\sigma$, with
dimensions of $(length)^{-2}$, is the string tension and $i$ goes
over the transverse degrees of freedom.
Let $\kappa\in [0,r] $ be the coordinate along the string. Then 
the configuration and transverse momenta of the string can be represented as
\begin{eqnarray}
x_i(\kappa)=\sum_{n=1}^{\infty} x^n_i \; \sin(\frac{n\pi\kappa}{r}),\\
{\cal P}_i(\kappa)=\sum_{n=1}^{\infty} {\cal P}^n_i \; \sin(\frac{n\pi\kappa}{r}).
\end{eqnarray}
The Hamiltonian is now given by
\begin{equation}
{\cal H}=\frac{\pi}{4r}\sum_{n=1}^{\infty}\left [ \frac{({\cal P}^n_i)^2}{\sigma}
+n^2 \sigma (x^n_i)^2 \right ].
\end{equation}
To go to the number operator basis, let us now define the creation and 
annihilation operators such that
\begin{equation}
x^n_i=\frac{1}{\sqrt{n\sigma}}(a^{\dagger n}_i + a^n_i ) \; 
{\textnormal{and}} \;
{\cal P}^n_i=i\sqrt{n\sigma}(a^{\dagger n}_i - a^n_i),
\end{equation}
with $[a^{\dagger n}_i,a^m_j]=\delta_{ij}\delta^{nm}$.
In terms of the creation and annihilation operators, the Hamiltonian can be 
formally written as 
\begin{equation}
{\cal H}=\sum_{n=1}^{\infty} \frac{n\pi}{r}(a^{\dagger n}_i a^n_i +
\frac{(d-2)}{2}).
\end{equation}
In this expression, the second term is divergent and we use zeta function 
regularisation for it. Using $\zeta (-1)=-\frac{1}{12}$, we can rewrite 
the Hamiltonian as 
\begin{equation}\label{eH}
{\cal H}=\left (\sum_{n=1}^{\infty} \frac{n\pi}{r}a^{\dagger n}_i a^n_i 
\right ) - \frac{\pi}{24 r} (d-2)
\end{equation}
where $d$ is the number of space-time dimensions. Thus from the zero point 
energy we obtain the universal
$1/r$ contribution to the potential at large $r$. For a similar derivation
see \cite{BN}. From equation (\ref{eH}), we
 read off the energy difference between successive states to be $\pi/r$.
These are two very important predictions of the string picture and we will 
confront both of them with our data.

In 2+1 dimensions, we have only one set of oscillators since there 
is only one transverse dimension and the eigenstates of the Hamiltonian 
fall under four different channels distinguished by their behaviour under
the discrete transformations of parity and charge conjugation. In 
contrast, in 3+1 dimensions there are two transverse directions and hence 
two independent sets of oscillators. This brings in an additional angular 
momentum quantum number given the usual 2-d harmonic oscillator algebra.

We will look at 2+1 dimensions and there, since we have only one transverse 
direction, we drop the index $i$.
To classify our states, we now determine the behaviour of $a^n$ 
and $a^{\dagger n}$ under parity and charge conjugation.
For this it is sufficient to note that parity implies 
$x(\kappa)\longrightarrow - x(\kappa)$ and charge 
conjugation takes $x(\kappa)\longrightarrow x(\pi-\kappa)$.
From this definition it is easy to see that under parity
$\{a^n, a^{\dagger n}\}\longrightarrow \{-a^n, 
-a^{\dagger n}\}$ and under charge conjugation $\{a^n, 
a^{\dagger n}\}\longrightarrow \{(-1)^{n+1}a^n, 
(-1)^{n+1}a^{\dagger n}\}$.
Below we list the first few string states along with their $C$
 and $P$ quantum numbers.

\begin{center}
\vspace*{5mm}
\begin{tabular}{r|r|r|r}
 \hline 
+ + \hfill & + $-$ \hfill & $-$ $-$ & $-$ + \hfill \\
\hline 
$|0\rangle$ & $a^{\dagger}_1|0\rangle$ & $a^{\dagger}_2|0\rangle$ & 
$a^{\dagger}_2a^{\dagger}_1|0\rangle$ \\
&&&\\
$(a^{\dagger}_1)^2|0\rangle$ & $(a^{\dagger}_1)^3|0\rangle$ & 
$(a^{\dagger}_1)^2 a^{\dagger}_2 |0\rangle$ & $(a^{\dagger}_1)^3 a^{\dagger}_2 
|0\rangle$\\ 
& $a^{\dagger}_3|0\rangle$ & $a^{\dagger}_4 |0\rangle$ & $a^{\dagger}_1 a^{\dagger}_4 
|0\rangle$ \\
&&& $a^{\dagger}_2 a^{\dagger}_3 |0\rangle$ \\
&&& \\
$(a^{\dagger}_1)^4|0\rangle$ &&& \\
$a^{\dagger}_1 a^{\dagger}_3 |0\rangle$ &&& \\
$(a^{\dagger}_2)^2|0\rangle$ &&& \\
\hline 
\end{tabular}
\vspace*{5mm}
\end{center}

Here we will look at the ground states in each channel as well
as the first excited state in the $\{+ +\} $ channel.

\section{Simulation Details}

In this work we perform three different kinds of simulations.
First we measure Polyakov loop correlation functions. Apart from giving
accurate information about the ground state, this also helps us set the 
scale on the lattice. Secondly, to look at the ground states in various channels, 
we perform simulations with sources belonging to different channels.
Finally to investigate excited states in the same channel we use sources of the 
same shape 
but create correlation matrices by using different smearing parameters.
An outline of these procedures is given below.
 
Most of our calculations are carried out at $\beta=5$ on a $24^3$ lattice. 
However we also perform some 
measurements at larger $\beta$ and lattice volume.

\subsection{Parameters of the algorithm}

We work with three couplings viz. $\beta=5$, $7.5$ and $10$.
The lattice volumes are $24^3$, $36^3$ and $48^3$ respectively. At $\beta=5$,
we also measure Polyakov loops on a lattice of extent $24$ in the space direction
but $8$ in the time direction.

For our simulations we use the L\"uscher-Weisz multi-level algorithm.
Since this algorithm is relatively new, we list some of the parameters
that we use with this algorithm. We restrict ourselves to one level of
averaging and estimate the average of the product of two 2-link operators
over a number of sub-lattice updates \footnote{For a definition of these 
terms see \cite{multi}}. (At $\beta=10$ we use a product of four 2-link 
operators instead of two). For each sweep of heat-bath we use three sweeps of
over-relaxation. In table 
\ref{param} we list the 
number of sub-lattice updates (iupd) we use for various quantities.

\begin{table}
\begin{center}
\begin{tabular}{c|c|c|c}
\multicolumn{4}{c}{Polyakov Loop} \\
\hline
$\beta$ & 5 & 7.5 & 10 \\
\hline
iupd & 1600 & 1600 & 3200 \\
\hline 
\end{tabular}
\begin{tabular}{c|c|c|c|c|c}
\multicolumn{6}{c}{Wilson Loop (all $\beta$ values)} \\
\hline
$T$ & 2 & 4 & 6 & 8 & 12 \\
\hline
iupd & 100 & 200 & 300 & 400 & 600 \\
\hline
\end{tabular}
\caption{\label{param} Parameters used in the algorithm.}
\end{center}
\end{table}

At $\beta=5$ these values are close to optimal. For the Polyakov loops at
$\beta=7.5$, the optimal value for the number of sub-lattice updates is probably
more than what we used. 
$\beta=10$  required a different averaging scheme. We had to use a product of 
four 2-link operators instead of two to obtain  
the kind of error reduction as at the lower values of $\beta$.

\subsection{Setting the scale}

We use the string tension to set the scale on the lattice.
To obtain the string tension, we begin by measuring the Polyakov loop 
correlation functions for various separations. The Polyakov loop correlator,
at a distance $r$ can be represented as
\begin{equation}\label{bcoeff}
\langle P(r,T)P(0,T) \rangle = \sum_{i=0}^{\infty} b_i \; \exp[-V_i(r)T]
\end{equation}
where $b_i$'s are integers \cite{LW1}. At large $r$ where the string picture is thought
 to be valid, $V(r)=V_0(r)$ is given by  
$\sigma r \: + {\tilde V}\:  + c/r + \ldots $ where ${\tilde V}$ is a constant 
depending only on $\beta$, $\sigma$ is the string tension, $c$ is the universal 
constant mentioned above and the dots represent terms with higher inverse 
powers of $r$. In three dimensions $c=-\pi/24$. 

From correlation functions, we obtain the
potential $V_0(r)$ by 
\begin{equation}\label{polen}
V(r)=-\lim_{T\rightarrow\infty}\frac{1}{T}\log\langle P(r,T)P(0,T) \rangle 
\end{equation}
Then we define the force $F(r)$ by the symmetric difference of the potential
 i.e. $F(r)=[V(r+1)-V(r-1)]/2$. We also compute $c$ at every value of $r$ by 
defining $c(r)=[V(r+1)+V(r-1)-2 V(r)]r^3/2$ . These are tabulated in table 
\ref{force}. 
To obtain the string tension, we plot $F(r)$ against $1/r^2$ in Fig. 1. 
Then the intercept gives us the value of 
the string tension $(a^2\sigma )$ and the asymptotic value of the slope gives us 
the constant $c$.

\begin{figure}[htb]
\begin{center}
\mbox{\epsfig{file=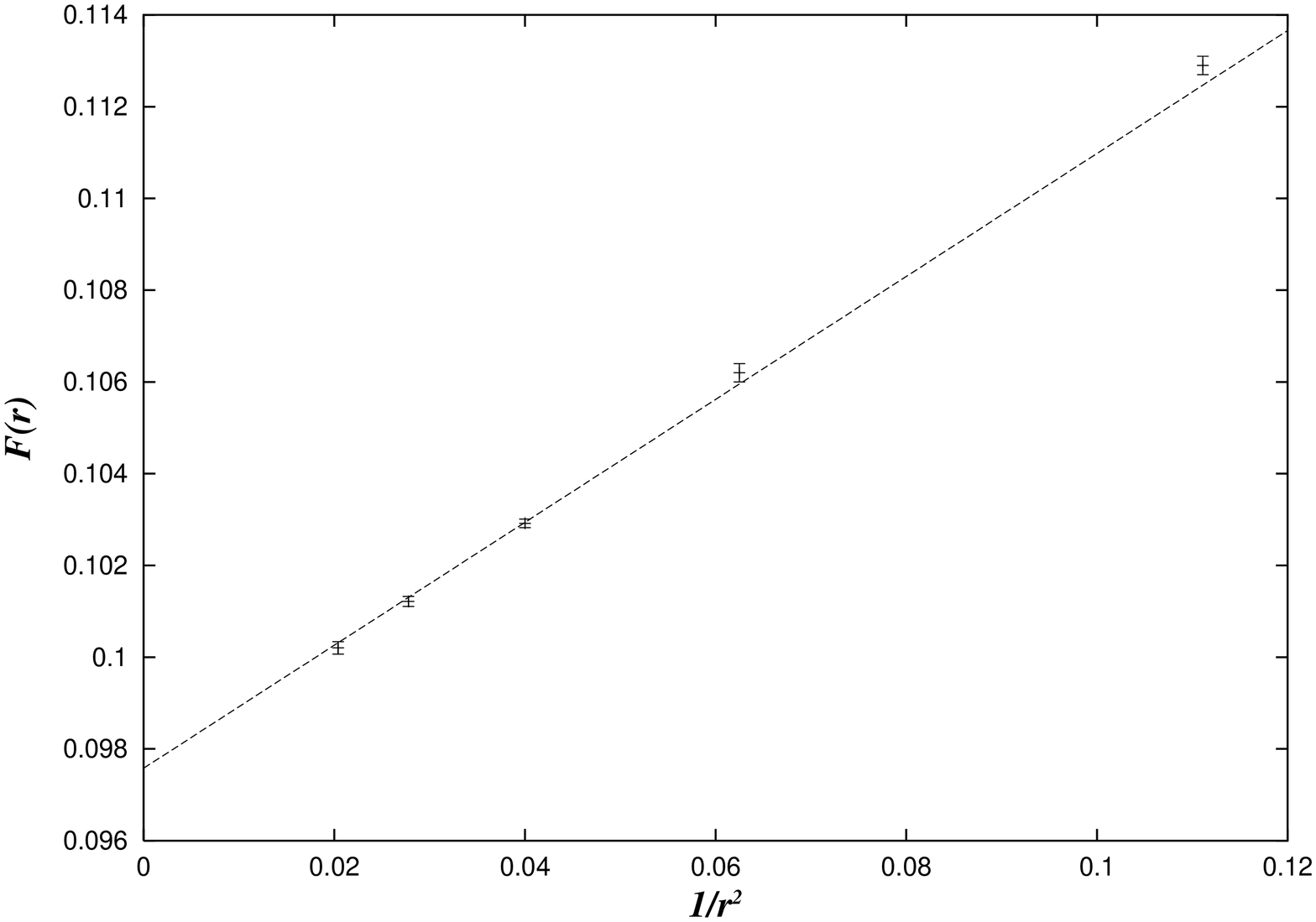,width=10truecm,angle=-90}}{ \\ Fig. 1. 
The force $F(r)$ extracted from the Polyakov loop vs $1/r^2$ 
for lattice size 24$^3$. $\beta=5$.}
\end{center}
\end{figure}

The only problem with the above analysis is that the string behaviour is
expected only at large distances and since time and computational resources
prevent us from going to too large distances, the quantities determined from 
the plots have a significant admixture of short distance effects. Therefore
it is preferable to use the locally determined $c(r)$'s for as large $r$'s
as possible to set the slope and then determine the intercept to get the 
string tension. For the 24$^3$ lattice we were able to use $c(r)$ to set the
slope. However for the other two cases $c(r)$ was not sufficiently accurately 
determined to be of use. In those cases both the parameters were obtained from 
the graph. In table \ref{scale1} we tabulate $a\sqrt \sigma $ and the Sommer 
scale $r_0$ \cite{Sommer} defined by $r_0^2F(r_0)=1.65$.
In the table we quote $\sqrt \sigma r_0$. The scale can be set explicitly in 
different ways, for example, we use $\sqrt \sigma = (0.5 \;fm )^{-1}$.

At $\beta =5$, our result for $a\sqrt \sigma $ compares nicely with Teper's
\cite{Teper} value of 0.3129 (20). We also look at how $c(r)$ compares with
what we obtain from perturbation theory. From the two loop static quark 
potential $V(r)$ in \cite{static} we obtain the perturbative value of $c(r)$ 
at $\beta =5$ as
\begin{equation}\label{pe}
c(r)=\frac{r^3}{2}\frac{\partial^2 V(r)}{\partial r^2} 
=-\frac{3r}{4\pi\beta}=-\frac{3r}{20\pi}
\end{equation}
In Fig. 2 we plot this perturbative value along with the values of $c(r)$
given in table \ref{scale1}.
\begin{figure}[htb]
\begin{center}
\mbox{\epsfig{file=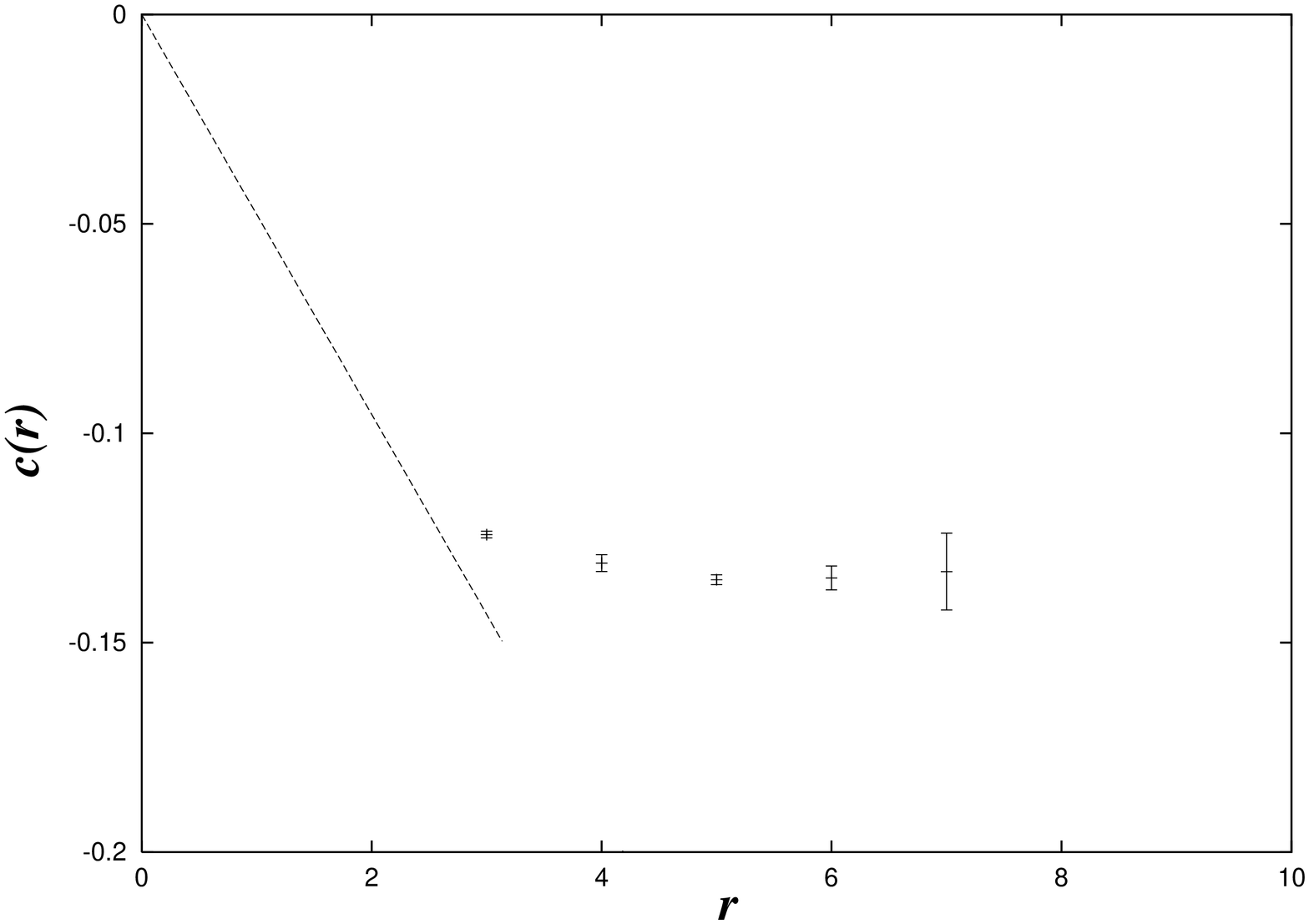,width=12truecm,angle=0}}{ \\ Fig. 2.
$c(r)$ at $\beta=5$. The curve is 2 loop perturbation theory.}
\end{center}
\end{figure}
This figure shows the transition from perturbative to non-perturbative 
behaviour for $c$. This is interesting because although the short distance
behaviour depends on the gauge theory, the long distance string picture 
gives the same value for all gauge theories. For similar results in SU(3),
see \cite{LW1}.

\subsection{Correlation matrices}

 The advantage of the Wilson loop is 
that it also offers the possibility to study the higher
 excited states.  This can be done by using sources which couple preferentially to 
the required state. If the optimal sources are unknown, this can be achieved by 
forming correlation matrices $C(r,T)$ among various wave functions. 
Upon diagonalisation, the linear combination of the wave functions which give the 
eigenvectors of $C(r,T)$ are the required sources.
The matrix element $C_{ij}(r,T)$ is nothing but a Wilson loop of extent $T$ in the 
time direction, with $source_i$ at one end and $source_j$ at the other. The 
sources are paths between points separated by distance $r$.

To explore the four different channels, we use a basis of four paths shown in
 Fig. 3.
\begin{figure}[htb]
\begin{center}
\mbox{\epsfig{file=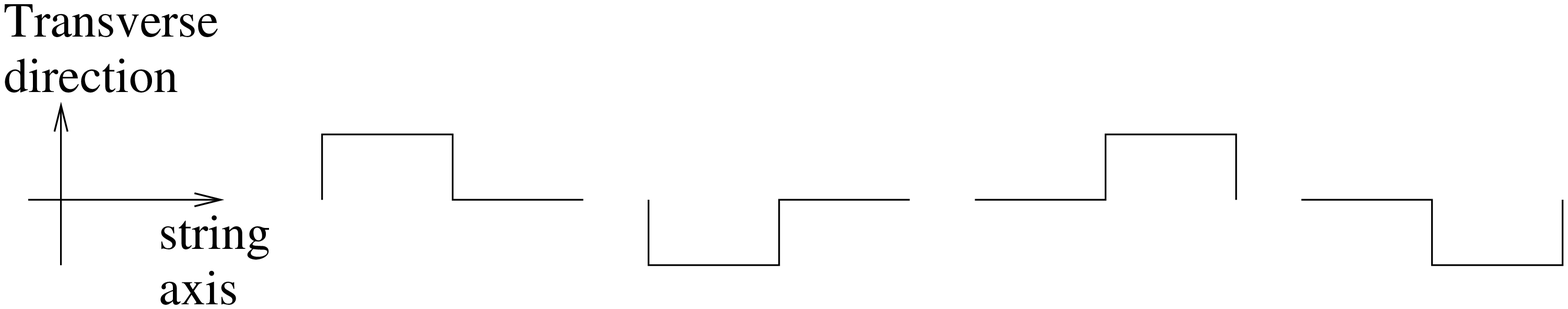,width=10truecm,angle=0}}
{\\ Fig. 3. The basis elements of the correlation matrix.}
\end{center}
\end{figure}
where the staples are of length $\lfloor r/2 \rfloor $.
 
These paths can be combined into orthogonal
channels invariant under the action of $C$ and $P$ as shown in Fig. 4.
\begin{figure}[htb]
\begin{center}
\mbox{\epsfig{file=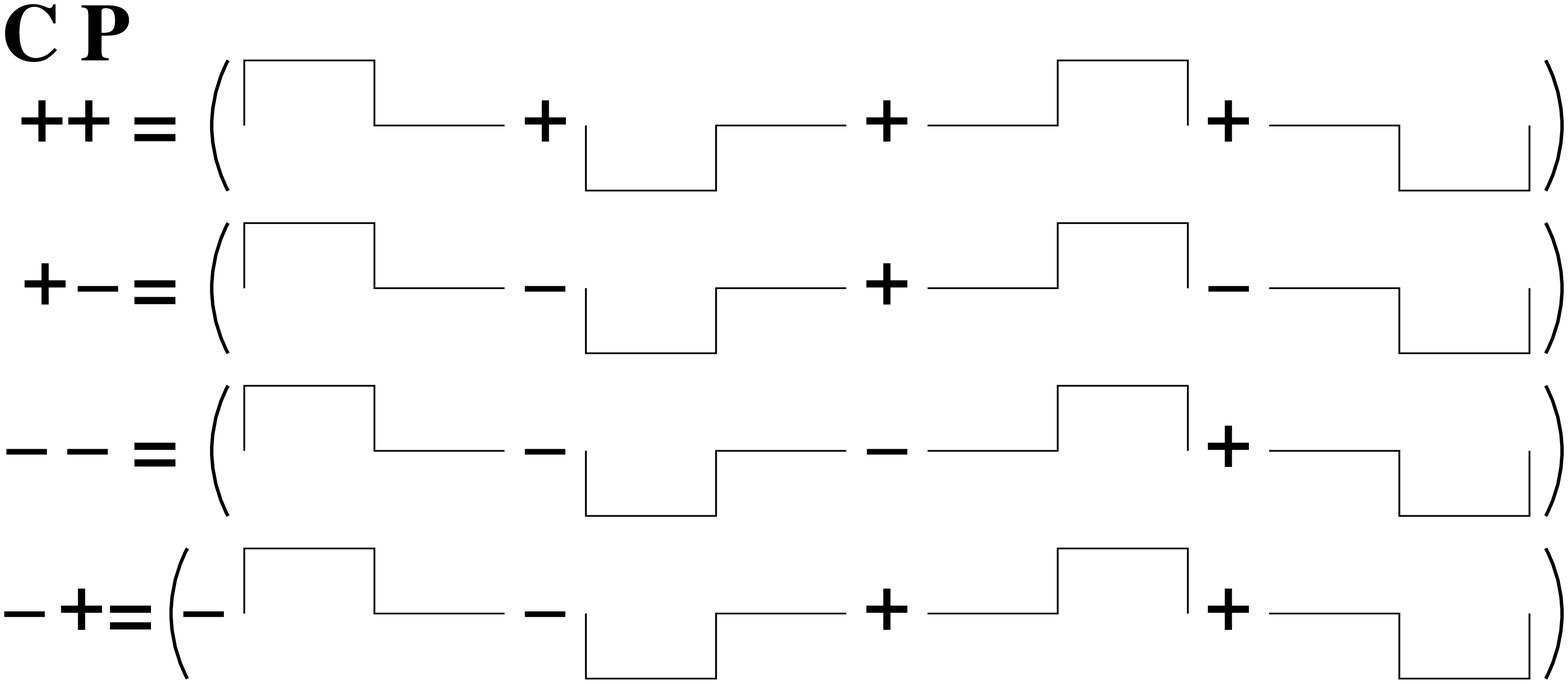,width=10truecm,angle=0}}
{\\ Fig. 4. The eigenstates of the correlation matrix.}
\end{center}
\end{figure}

To look at excited states in the $\{++\}$ channel, we take the straight
path but create different sources by using different smearing parameters.
Thus in this case our sources are defined as 
\begin{equation}
source_i= {\cal P}\left( \prod_{k=1}^r(U + \gamma_i \sum S )\right )
\end{equation}
where $\gamma_i \in \{0.1, 0.3, 0.5\}$ and ${\cal P}$ denotes projection back to 
SU(2). $U$ is the original link and $S$ are the space-like staples.
This procedure can be applied recursively and we do four levels of smearing.
While this may not be an optimal choice for the sources, it was sufficient
for our purposes.  Another advantage of smearing is that it reduces
fluctuation of the sources.
So even when we use the different paths to probe the different 
channels, we use one level of smearing with $\gamma=0.5$.

The eigenvalues of the correlation matrix, for large time separations, are 
related to the 
energies of the various states of the sources. We find that these matrices are 
always dominated by the ground state and that the higher excited states are 
suppressed by several orders of magnitude.

\section{Results}
\subsection{Polyakov loops}
From the Polyakov loop correlators, we get energies of 
the ground state and the first excited state. At $\beta=5$, the 
ground state is obtained from a $24^3$ lattice and the first excited state from
a combination of $24^3$ and $8\times 24^2$ lattices.

The energy is given by equation (\ref{polen}). The corrections to the energies 
are due to finite volume and mixing with higher energy states. Both of these 
can be quite reliably
estimated. First we look at the leading order finite volume correction. 
Neglecting the higher energy states, the measured energy of a state is 
\begin{equation}\label{polcorr1}
-\frac{1}{T}\log (\langle PP \rangle (r)) =  
V(r) -\frac{1}{T}\log\left (1+e^{-(L-2r)(\sigma -c/r(L-r))T}\right ).
\end{equation}
Here $V(r)$ is an improved estimate of the energy and the second term is the 
finite volume 
correction. We will call the argument of the $\log$ ($1+{\cal V}(r))$. $L$ is the 
extent of the lattice in the space direction which here is $24$. 

Next we come to the correction due to the presence of higher energy states.
We will compute the correction to the first excited state. 
There are two states which contribute to the second excited state of
the Polyakov loop. They correspond to $a_1^{\dagger\;2}| 0\rangle $ and the
$a_2^{\dagger}| 0\rangle $. However we have explicit measurements for both
these states. Thus our correction factor is given by 
\begin{equation}\label{polcorr2}
{\cal Q}(r,T)=e^{-TE_1^{++}(r)}+e^{-TE_0^{--}(r)}.
\end{equation}
We estimate further higher order corrections to be smaller than our statistical
errors.

The measured energy obtained from the $24^3$ lattice is 
overwhelmingly dominated by the ground
state with the finite volume effects and effects from the higher states being
smaller than our statistical errors. In contrast the value measured from the 
$8\times 24^2$ lattice contains a significant contribution from the first 
excited state.
To extract this state we subtract the ground state contribution, as 
obtained from the $24^3$ lattice, from the $\langle PP \rangle$ expectation 
value. From equations (\ref{polcorr1}) and (\ref{polcorr2}) we also expect 
 finite volume effects and contamination due to higher excited states for $V(r)$
from the $8\times 24^2$ lattice. 

Taking all these into account, our formula for the first excited state from the 
Polyakov loop is given by
\begin{equation}
E_1(r)=-\frac{1}{8}\log \left ( \frac{\langle PP \rangle (r)}{1+ {\cal V}(r)} 
-e^{-8 E_0(r)}-{\cal Q}(r,8) \right )
\end{equation}
where ${\cal V}$ is the leading order finite volume correction and ${\cal Q}$ is 
the correction due to higher states. 
Our results are tabulated in table \ref{polya}.

In the preceding analysis we have assumed that the coefficients $b_0$ and 
$b_1$ given in equation (\ref{bcoeff}) are unity as predicted by the free 
theory. Agreement of the energies with 
the ones determined from the Wilson loops bears out this assumption. 
There are departures from the free theory predictions too. Free 
theory predicts $b_2=2$ which implies a degenerate $E_1^{++}$ and $E_0^{--}$ state. 
However explicit measurement shows that this degeneracy is lifted and these states
occur separately with coefficients unity. Since the 
coefficients can only change by integers, any other value for all the four $b$'s 
is effectively ruled out.

\subsection{Wilson Loops}

We first use the Wilson loop to look at the
ground states of the four different channels. In this case since we
know the symmetries of the invariant sub-spaces under ${\bf C}$ and ${\bf P}$, we 
directly take the appropriate combinations of the matrix 
elements to diagonalise the correlation matrix.
The energies are obtained by taking the ratio of the eigenvalues at
fixed $r$ but different $T$'s. Thus they are given by 
\begin{equation}\label{wilson1}
E=-\frac{1}{T_2-T_1}\log\frac{\lambda(r,T_2)}{\lambda(r,T_1)} 
\end{equation}
where $T_2 > T_1$ and $\lambda(r,T)$ is an eigenvalue of the 
correlation matrix $C(r,T)$  
This however is just a naive estimate of the energies. Actually there are
contributions from the higher states. That is why it is desirable to
choose as large a value of $T$ as possible \cite{LW}. To try to take 
this into account let us expand $\lambda$'s as 
\begin{equation} 
\lambda (r,T)=\alpha_1 e^{-E_0(r) T}+\alpha_2 e^{-E_1(r) T}+\ldots 
\end{equation}
where $E_0$ is the ground state in a given channel and $E_1$, $E_2$
 etc. are the higher energy states in the same channel. We will estimate
the correction to the ground state by taking into account the contribution
only due to the next excited state.
At a fixed value of $r$, let 
\begin{eqnarray}
\lambda (T_1)&=&\alpha_1 e^{-E_0 T_1}\left (1+\frac{\alpha_2}{\alpha_1} 
e^{-\delta T_1} \right ), \\
\lambda (T_2)&=&\alpha_1 e^{-E_0 T_2}\left (1+\frac{\alpha_2}{\alpha_1}
e^{-\delta T_2} \right )
\end{eqnarray}
where $\delta=(E_1-E_0)$ to leading order.
Then the correction to the energy is given by
\begin{equation}\label{wilsoncorr}
-\frac{1}{T_2-T_1}\log\frac{\lambda(T_2)}{\lambda(T_1)} =
{\bar E} + \frac{1}{T_2-T_1}\left [ \frac{\alpha_2}{\alpha_1}
e^{-\delta T_1}\left ( 1-e^{-\delta (T_2-T_1)}\right ) \right ].
\end{equation}
Here the left hand side of the equation is the measured energy
and ${\bar E}$ is the improved estimate.
As seen from this expression the correction has an exponential
dependence on $T_1$ and a linear dependence on $T_2-T_1$. Both these
trends are seen in the data. 

We concentrate mostly on the $24^3$ lattice at $\beta=5$. 
On this lattice, we measure correlation matrices with $T$ extent's
of 2, 4, 6 and 8 respectively. We extract the energies by two procedures.
The first one is to do a fit to the form $\alpha \; \exp(-Et)$. This gives 
a naive estimate and these results are tabulated in table \ref{naive}.
In the second case we compute the energy from the six different combinations
of $T_1$ and $T_2$ and then fit it to the form given in equation
(\ref{wilsoncorr}) with ${\bar E}$
, $\frac{\alpha_2}{\alpha_1}$ and $\delta$ as the fit parameters. 
However this requires accurate data for stable fits and we were only able
to do this with confidence for the first two energy states. Our 
statistics for $T=8$ was not good enough for the higher states. Nevertheless
 at $\beta=5$, we were able to obtain an extrapolation for the third state
using only the $T=2,4$ and $6$ data. For the fourth state even that 
was not possible. All these values are given in table \ref{extra24}. 

From the data we estimate that the error in replacing $\log(1+ 
\frac{\alpha_2}{\alpha_1}
\exp(-\delta T_1)) $ by $\frac{\alpha_2}{\alpha_1}\exp(-\delta T_1) $ in 
equation (\ref{wilsoncorr}) is 
well below our statistical errors for the first two states. We also checked 
that $\bar{E}$
was within our error bars even with a 20\% change in the value of $\delta$. The
most stringent condition on the fits come from the value of 
$\alpha_2/\alpha_1$.
The maximal value of $\alpha_2/\alpha_1$ is the ratio of the degeneracy of the
states between the level considered and the next level. Any data set             
which violated this criterion, was considered not accurate enough to do a 
stable extrapolation to infinite $T$ and has not been used.

\begin{figure}[htb]
\begin{center}
\mbox{\epsfig{file=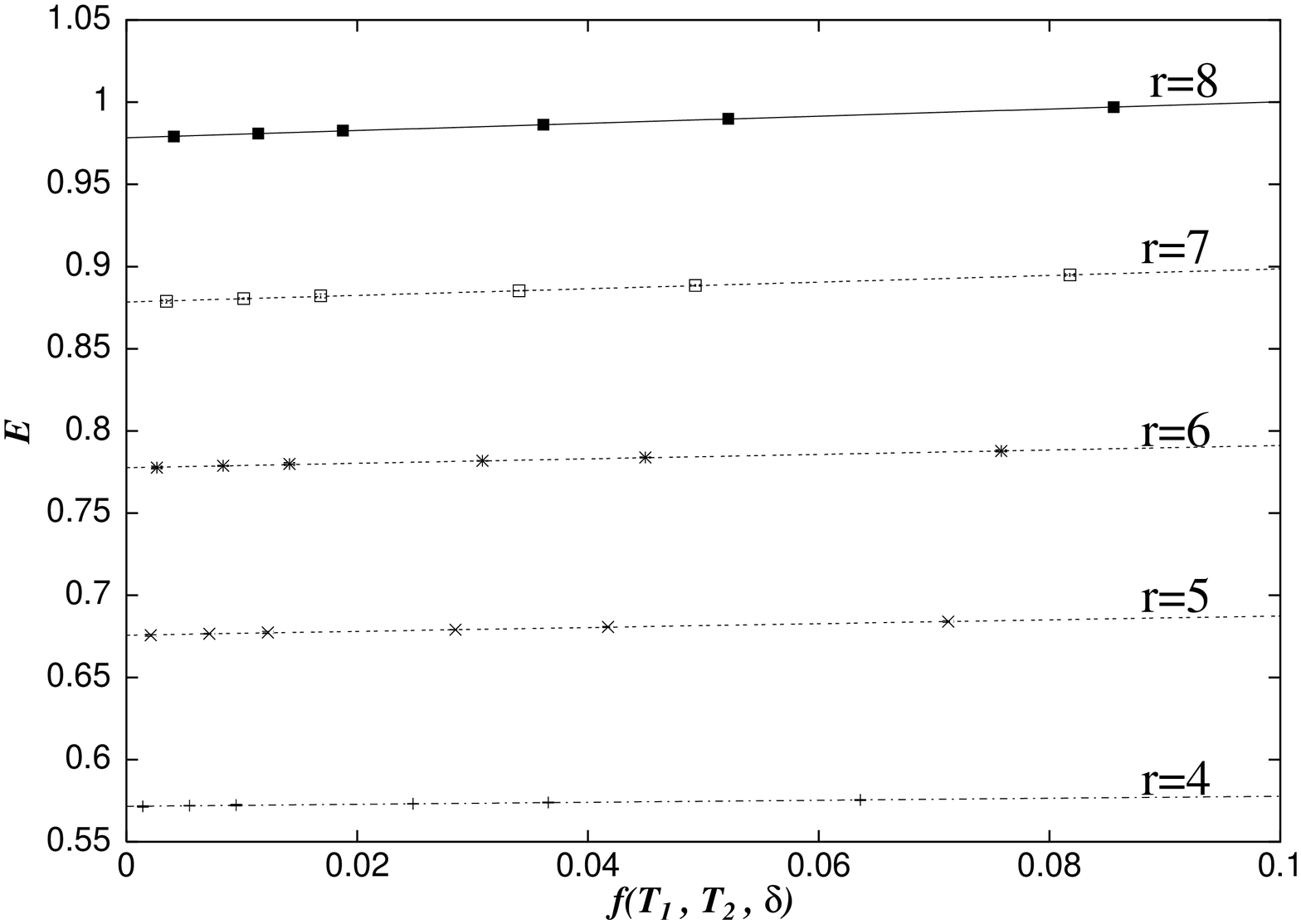,width=12truecm,angle=0}}
{\\ Fig. 5. Extrapolation of energies for ground state of \{++\}
channel. $\beta=5$.}
\end{center}
\end{figure}

For the \{++\} channel we have the explicit values for $E_0^{++}$
and $E_1^{++}$. This information
can be used to compute the second term in equation (\ref{wilsoncorr}) with 
$\frac{\alpha_2}{\alpha_1}$ and ${\bar E}$ as unknown parameters. The 
resulting data can be fitted to a form $ax+b$ to obtain ${\bar E}$ and 
$\frac{\alpha_2}{\alpha_1}$. These values are shown against $E_0^{*++}$ in 
table \ref{extra24}. As seen from the data, the error bars here are much larger 
than the ones on $E_0^{++}$. This is because the estimated energy difference 
between the two 
states is not constant, but varies with varying $T_1$ and $T_2$. We took
the mean value of this difference for $\delta$. That is a good approximation
for small $r$ where the difference is small, but becomes worse as
$r$ increases. The uncertainty reported in the values of $E_0^{*++}$ comes 
from the difference of the
fitted values if one uses the extreme values of $\delta$ instead of the mean.   
In Fig. 5 we plot the extrapolations for the $E_0^{*++}$ values.

In this plot the x-axis is the function $f(T_1, T_2, \delta)$ given by
\begin{equation}
f(T_1, T_2, \delta)=\frac{1}{T_2-T_1}\left [
e^{-\delta T_1}\left ( 1-e^{-\delta (T_2-T_1)}\right ) \right ] 
\end{equation}
The symbols $+$, $\times$, $\ast$, $\Box$ and $\blacksquare$ correspond to
$r$ values of 4,5,6,7 and 8 respectively. As seen from the plots,
the slope of the fitted lines increase with increasing $r$. 

To illustrate the importance of these corrections, we plot the energy difference
between the ground state and excited states at $\beta=5$ in Fig. 6.
\begin{figure}[htb]
\begin{center}
\mbox{\epsfig{file=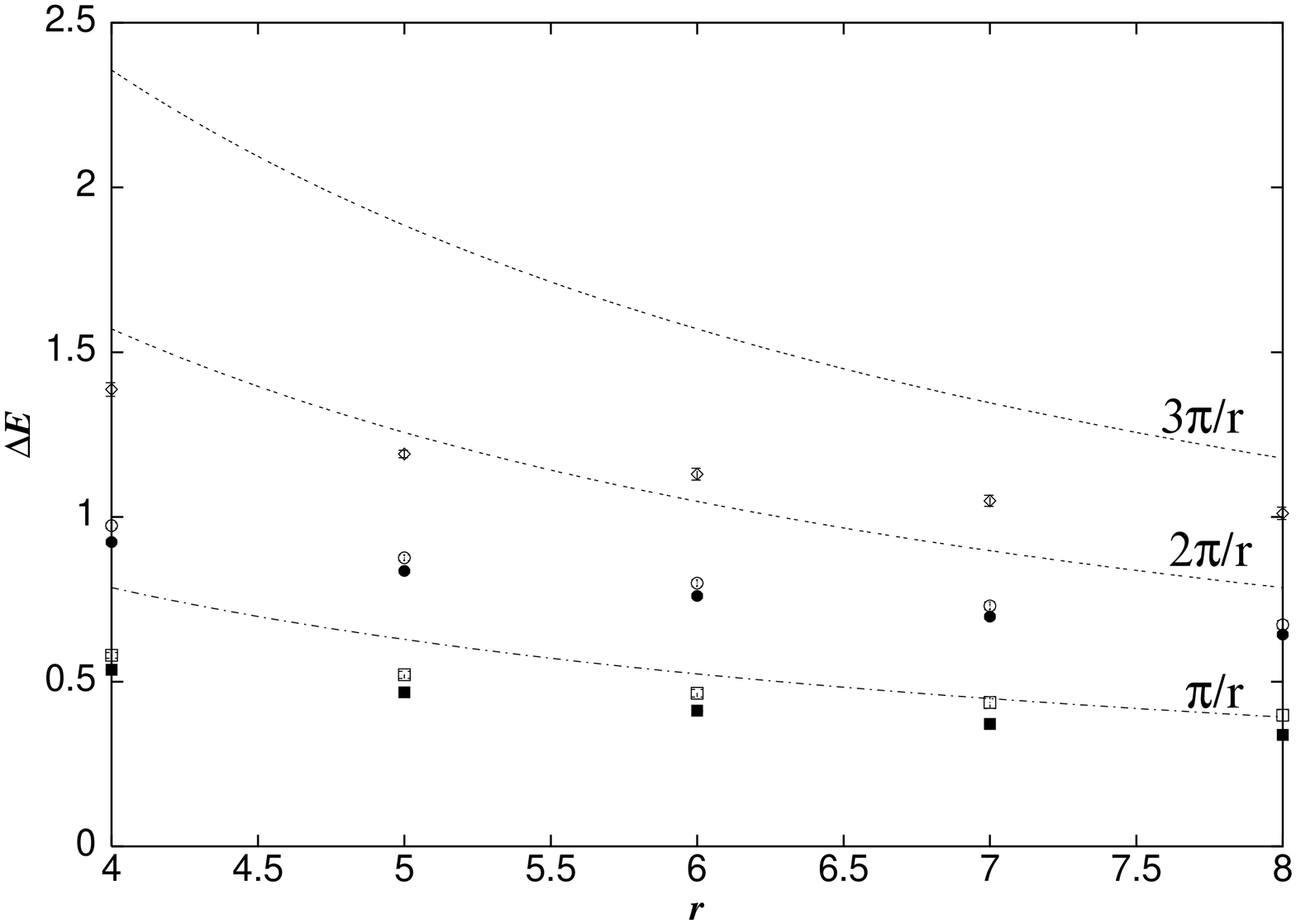,width=12truecm,angle=0}}
{\\ Fig. 6. Energy differences at $\beta=5$. The open symbols are naive energy
differences and the filled ones are the extrapolated ones.}
\end{center}
\end{figure} 
In this figure the open symbols $\Box$, $\circ$ and $\diamond$ denote the naive 
energy differences while filled symbols $\blacksquare$ and $\bullet$ are the ones 
obtained after performing  the
finite $T$ corrections. The curves are the expected energy differences of $\pi/r$,
$2\pi/r$ and $3\pi/r$. The uncorrected data set seems to contradict the string 
prediction as the energy difference between the ground state and the first excited
state tends to become more than the string value at large $r$. However the corrected
data shows no such trend but for large $r$ seems to approach a value close to the one 
predicted by the string theory. 

This behaviour persists even at higher values of $\beta$. In Fig. 7
we plot  
the energy difference between the first and the second states at $\beta =10$.
\begin{figure}[htb]
\begin{center}
\mbox{\epsfig{file=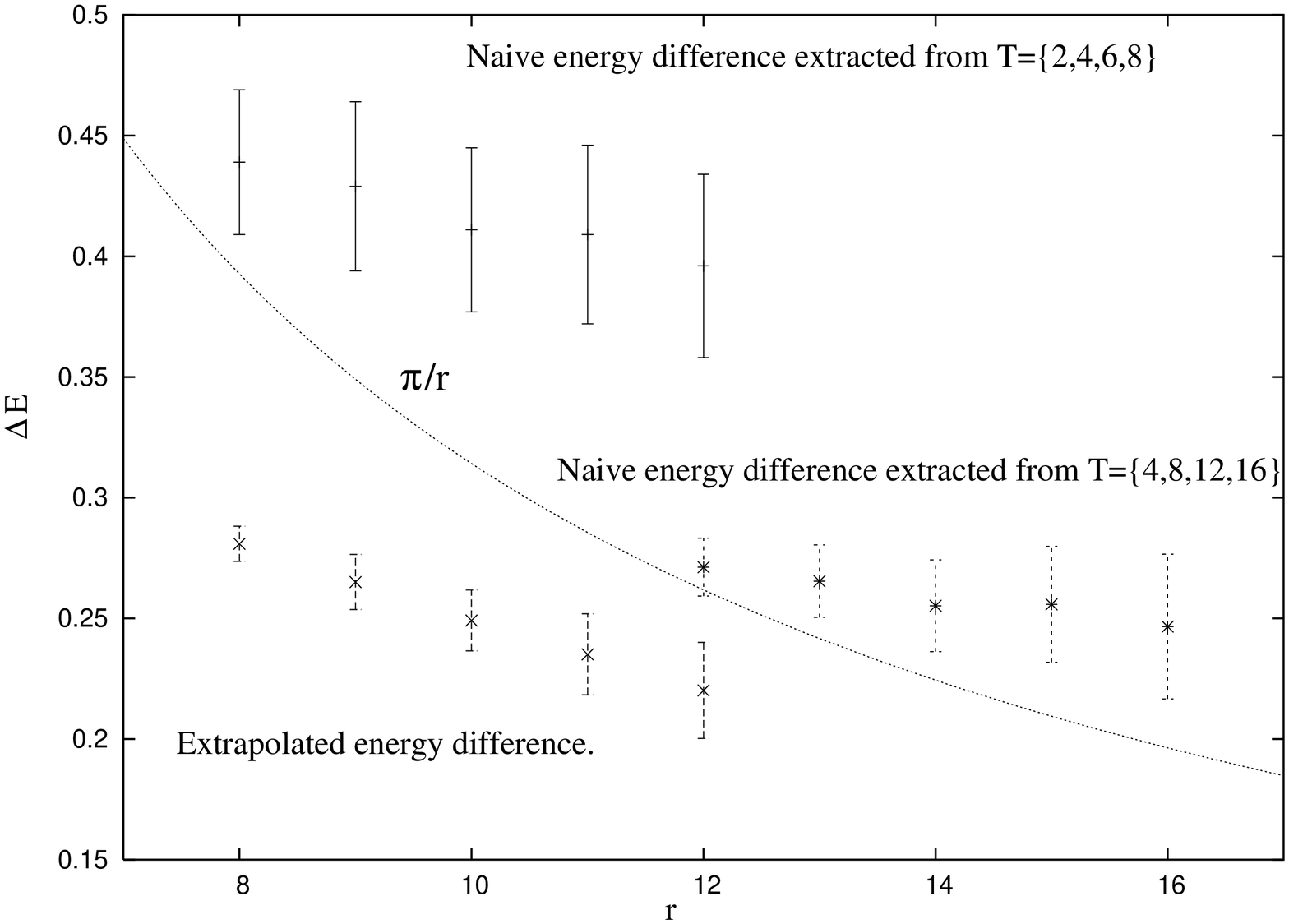,width=12truecm,angle=0}}
{\\ Fig. 7. Energy differences at $\beta=10$. This figure illustrates how the 
naive energy difference changes depending on the time extent of the correlation 
matrices. The lowest set of points show the energy difference obtained by 
doing the extrapolation to $T=\infty$ as described in the text. The dotted 
curve is the $\pi/r$ curve predicted by the free theory.}
\end{center}
\end{figure}
In this figure the topmost set of points are the naive energy differences 
between the states $E_0^{++}$ and $E_0^{+-}$. The lowest set of points are the 
energy differences obtained by doing the extrapolation to $T=\infty$ using the 
same data set. Finally the set in the middle shows the naive energy difference 
between the same states, but now determined from the correlation matrices with
larger $T$ extents. Unfortunately we do not have good enough data to do a 
$T=\infty$ extrapolation with this set. This difference clearly shows the 
importance of taking into account the corrections due to finite $T$ extent of 
the correlation matrices.

In \cite{asymm1} the energy difference between string states for SU(2) in three
dimensions is also presented. Sophisticated wave functions and improved anisotropic 
gauge actions are used in that study. However they do not report taking the finite $T$ 
corrections into account. They too see evidence of the energy difference crossing the 
$\pi/r$ curve predicted by naive string theory and point to the possibility of the 
existence of a massive QCD string. Qualitatively, the degeneracy and level ordering of 
 our uncorrected data is in agreement with that study at larger values of $r$ and we 
believe that if that study is extended to larger physical $T$'s or
the correction due to finite $T$ extents is taken into account this crossing would not
occur.

To look at the excited state in the \{++\} channel, we formed the correlation 
matrix using different
smearing parameters. In principle this matrix could be numerically 
diagonalised to
obtain the eigenvalues and the energies once again obtained by taking
the ratio of eigenvalues for different $T$'s.
 However diagonalisation of the correlation matrices showed that while the
eigenvector corresponding to the ground state was stable, 
the eigenvector corresponding to the first excited state varied  
quite a bit for different values of $T$. 

Since equation (\ref{wilson1}) assumes that the eigenvectors of both $C(T_1)$ and 
$C(T_2)$ are the same, we were unable to use this equation
 to extract the energies in this case. One way to force the eigenvectors to be 
equal is the diagonalise $C^{-1}(T_1)C(T_2)$, but even in this case 
it is not guaranteed that the eigenvectors are the same as we expect 
from the well determined single $C(T)$. 

We therefore followed a different 
procedure for the $E_1^{++}$ state. 
In this case, the correlation matrix was formed by using smeared 
sources as the basis states. Each element 
of this correlation matrix $C$  can be expanded as \cite{matel}
\begin{equation}
C_{ij}(r,T)=\sum_{\alpha} \beta_i^{\alpha}\beta_j^{\alpha}e^{-E_{\alpha}T}
\end{equation}
where $e^{-E_{\alpha}(r)T}$ is related to the eigenvalues of the transfer 
matrix and the product $\beta_i^{\alpha}\beta_j^{\alpha}$ is related to the 
overlap of the basis states between which this matrix element is taken.
We truncated this expression at the second term and did a fit to a
form $a_1e^{-E_0T}+a_2e^{-E_1T}$ (where $a_1\equiv\beta_i^0\beta_j^0$
and $a_2\equiv\beta_i^1\beta_j^1$) to extract $E_0$ and $E_1$. The results 
are tabulated in table \ref{samech}. The very nice agreement of the 
ground state energy assures us that the fit to $E_1^{++}$ is reliable.
Moreover since $C$ was a $3\times 3$ matrix, we did fits to several matrix 
elements to make sure that we obtained consistent values for the energy. Of course 
the coefficients $a_1$ and $a_2$ differed in each case.

Finally, in Fig. 8, we plot in a consolidated manner all the energy states at 
$\beta=5$ for $r$ values of 4,5,6,7 and 8.
\begin{figure}[htb]
\begin{center}
\mbox{\epsfig{file=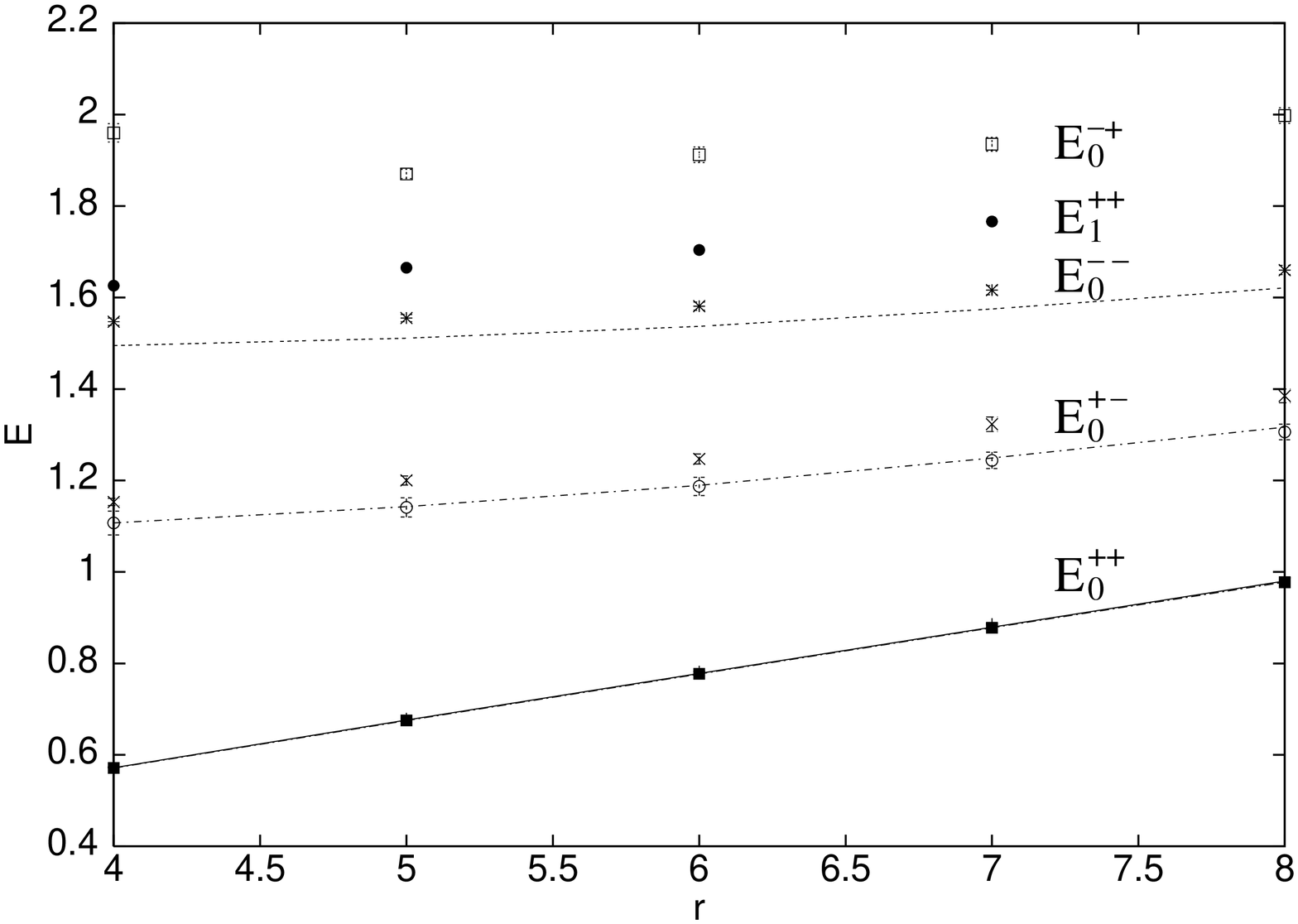,width=12truecm,angle=0}}
{\\ Fig. 8. The various energy states. Lattice size 24$^3$. $\beta=5$.}
\end{center}
\end{figure}
In this figure, the symbols $+$, $\times$, $\ast$ and $\Box$ correspond to 
the naive ground state energies of the four different channels \{++\}, 
\{+$-$\}, \{$--$\} and \{$-$+\}. The $\blacksquare$ corresponds to the 
ground state of the Polyakov loop and the line through it is the 
extrapolated ground state of the \{++\} channel. The $\circ$ corresponds
to the corrected first excited state from the Polyakov loop and the 
line through that is the extrapolated ground state in the \{+$-$\}
channel.
The dashed line just below the $\ast$ indicates the extrapolated values
in the \{$--$\} channel.
Finally the $\bullet$ corresponds to the first excited state in the 
\{++\} channel. 

The difference between 
the various values for the ground state in the \{++\} channel are not 
visible on this scale and in the figure only the $\blacksquare$ is visible. 
However the split between the naive value and the
extrapolated value is clearly visible in both the \{+$-$\} and 
\{$--$\} state. 
Our observation that the effect of extrapolation is higher for larger $r$
is also evident here as the difference between the naive and the 
extrapolated values increase with increasing $r$.

We also want 
to point out that although the corrections due to finite volume and higher
states are small effects, without taking them into account the excellent
agreement we get between the Polyakov loop values and the extrapolated 
Wilson loop values would not occur.

Our results for the energies at $\beta=7.5$ and $\beta=10$ are given in 
tables \ref{naive36} - \ref{extra48}.

\section{Comparison at different $\beta$}

To get some idea of how quantities depend on the lattice spacing, 
we look at the energy difference
between the ground state and the first excited state. 
To compare the differences at different $\beta$'s we need to evaluate 
them at the same physical distance. We choose three such points,
1.05 $r_0$, 1.2 $r_0$ and 1.35 $r_0$. 
Our results are contained in table \ref{ediff}.

\begin{figure}[htb]
\begin{center}
\mbox{\epsfig{file=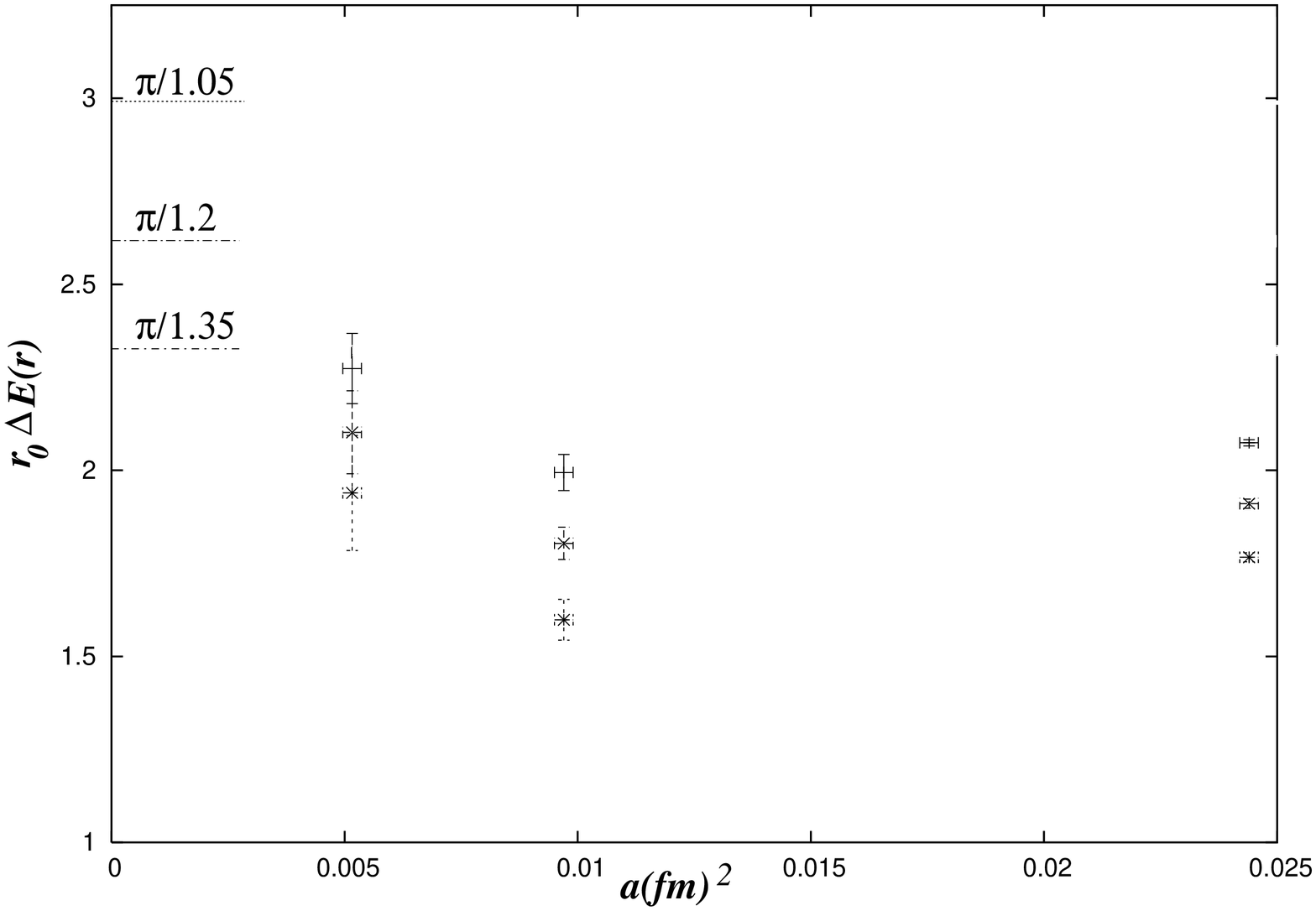,width=12truecm,angle=0}}
{\\ Fig. 9. The energy difference at different $\beta$.}
\end{center}
\end{figure}

In Fig. 9, the largest value of $a^2$ corresponds to 
$\beta=5$, and is believed to lie slightly outside the scaling region for SU(2) 
lattice
gauge theory in three dimensions. In the scaling region, 
according to the string prediction the energy differences should 
extrapolate to $\pi/1.05=2.992$, $\pi/1.2=2.618$ and $\pi/1.35=2.327$
 respectively. Since the approach to continuum seems to be non-monotonic, 
we are presently unable do a reliable extrapolation to the continuum. 

We also looked at the force and our findings are consistent with the 
expectation that the continuum limit is approached as ${\cal O}(a^2)$.
 
\section{Conclusion}

In this work we have looked at the potentials and forces between infinitely
heavy quarks and antiquarks and how they compare with predictions from
the effective hadronic string picture.

Using the L\"{u}scher - Weisz algorithm, we were able to go to previously 
inaccessible
distances and for the first time we were able to systematically probe the 
exponentially small corrections due to finite $T$. Reasonably large values of $T$ are 
required to see these effects and we were even able to extrapolate in some cases to 
infinite $T$. While these corrections are by themselves small, they are nevertheless 
absolutely crucial for comparing the data with the expected string spectrum.

Up to the distances we have measured, we have no reason to disbelieve the 
string 
prediction for the energy gap between the ground state and the first excited 
state. However it is necessary to use the extrapolated energy values to compute 
the difference. While the extrapolated data is not yet the free string 
value (which is supposed to hold at large $r$), it seems to approach it. However 
the naive data actually crosses this value and shows a rise with $r$. This 
crossing and rise with $r$ occurs at both $\beta=5$ as well as $10$. It is 
therefore very important to go to larger values of $r$ to verify whether these 
trends continue.
  
As we have mentioned before, we have not used any of the existing sophisticated 
machinery for the wave functions but concentrated on the systematic corrections
at larger $T$'s. The crucial point is that we have been able to take this  
correction into account order by 
order. We have only kept orders at which the corrections are larger 
than our statistical errors. Reasonably large values of $T$, allow us to see 
this effect on top of our
statistical uncertainties. The L\"uscher - Weisz algorithm lets us achieve 
this by making it possible to measure quantities accurately at the values of 
$T$ we have used.
However we expect that more accurate studies will 
require better wave functions as well, as the L\"{u}scher - Weisz algorithm 
alone is not enough to reduce the fluctuation of the sources.    

Finally it is clear that a more reliable continuum limit is necessary to confront 
the string predictions with lattice data and therefore studies at higher $\beta$ 
have to be undertaken.

\section{Acknowledgements}

The author would like to express his deep gratitude to Peter Weisz for innumerable
discussions and constant encouragement. Thanks are also due to Erhard Seiler, Ferenc
Niedermayer, Philippe de Forcrand and Amol Dighe for several useful suggestions. 
Finally the computations 
were carried out at the theory cluster at Max-Planck-Institute Munich. The author is
indebted to the institute for this facility. 

\appendix 
\section{Error Analysis}
 
We have two kinds of data. The directly measured quantities and the
derived quantities. The directly measured quantities are the matrix
elements obtained from measurement of the various Wilson loops
and the Polyakov loop correlation functions. The derived quantities are
the various energies which are determined from the matrix elements.
Below we outline how the errors were calculated in each case.
 
For all the Polyakov loop correlators and the ground states obtained
from the different path combinations, the errors were calculated using the
usual binned jackknife procedure. On the other hand, for the states in the 
same channel we used the naive errors since there was no
bin size for which the errors on all the matrix elements were maximized.
Also the naive and maximum errors differed only by about 15\%.

Let us now come to the derived quantities. 
While calculating the errors on the forces and $c(r)$, it is
possible to use the correlations between different values of $r$ to ones
advantage. To do so, $F(r)$ and $c(r)$ were calculated
individually for each measurement and then we used the jackknife analysis
on the various values of these quantities. However in this case too first
we had to bin the data so that $F(r)$ and $c(r)$ determined from the
bin averages satisfied all the known properties, which in this case are
$c(r)$ being positive and $F(r)$ decrease
with $r$ as can be seen from the convexity of the potential 
\cite{Seiler} \cite{convex}. Finally there are the various fitted
energy values. For the naive fits, the errors quoted are those calculated by the 
gnuplot fitting routine. For the extrapolated energies,
these errors were unrealistically small and in those cases we used
fits to ``central value + error" to get an estimate of the errors on the 
fitted values.

\newpage
\begin{landscape}
\begin{table}
\begin{tabular}{rll|rll|rll}
\hline
\multicolumn{3}{c}{ $\beta $ =5 } & \multicolumn{3}{c}{ $\beta $ =7.5 }& 
\multicolumn{3}{c}{ $\beta $ =10} \\
\hline
$r$ & \hspace*{5mm}$F(r)$ & \hspace*{5mm}$c(r)$ & $r$ & \hspace*{5mm}$F(r)$ & 
\hspace*{5mm}$c(r)$ & $r$ & \hspace*{5mm}$F(r)$ & $c(r)$ \\
\hline
3  &  0.1129 (2) & $-$0.1242 (8) & 5 & 0.04350 (8) & $-$0.110 (21) & 7 & 0.02318 (3) & $-$0.107 (1)\\
4  & 0.1062 (2) & $-$0.131 (2) & 6 & 0.0421 (1) & $-$0.119 (30) & 8 & 0.02266 (3) & $-$0.110 (1)\\
5  & 0.10320 (6) & $-$0.135 (1) & 7 & 0.0412 (1) & $-$0.109 (16) & 9 & 0.02224 (4) &$-$0.116 (2)\\
6  & 0.10167 (8) & $-$0.1345 (29) & 8 & 0.0407 (2) & $-$0.107 (70) & 10 & 0.02196 (4) &$-$0.118 (3)\\
7  & 0.10093 (9) & $-$0.133 (9) & 9 & 0.0395 (11) & \hspace*{5mm}- & 11 & 0.02175 (5) &$-$0.120 (4)\\
&&&&&&14 & 0.02133 (5) & $-$0.119 (17) \\
&&&&&&15 & 0.02125 (6) & $-$0.121 (27) \\
\hline
\end{tabular}
\caption{\label{force} $F(r)$ and $c(r)$ at different $\beta$ and $r$.
A `-' indicates stable measurement was not possible.}

\end{table}
\end{landscape}

\newpage

\begin{table}
\begin{center}
\begin{tabular}{r|c|l|l|l|l}
\hline
$\beta $ & lattice size & \hspace*{5mm}$c$ & $a\sqrt \sigma $ & $r_0/a$ & $\sqrt \sigma 
r_0$ \\
\hline
5.0 & 24$^3$ & 0.134 & 0.3124 (2) & 3.94 & 1.231 (1) \\
7.5 & 36$^3$ & 0.117 (3) & 0.1970 (2) & 6.285 & 1.238 (1) \\
10.0 & 48$^3$ & 0.133 (4) & 0.1437 (2) & 8.58 & 1.233 (2) \\
\hline
\end{tabular}
\caption{\label{scale1} String tension and Sommer scale.}
\end{center}
\end{table}

\begin{table}
\begin{center}
\begin{tabular}{clllll}
\hline
state & $r=4$ & $r=5$ & $r=6$ & $r=7$ & $r=8$ \\
\hline 
E$_0$ & 0.5716 (2) & 0.6755 (3) & 0.7773 (3) & 0.8778 (4) & 0.9775 (5) \\
E$_1$ & 1.107 (26) & 1.141 (21) & 1.187 (20) & 1.244 (18) & 1.306 (17) \\
\hline
\end{tabular}
\caption{\label{polya} Energies from Polyakov loop $\beta=5$ Lattice $24^3$.}
\end{center}
\end{table}

\begin{table}
\begin{center}
\begin{tabular}{clllll}
\hline
states & $r=4$ & $r=5$ & $r=6$ & $r=7$ & $r=8$ \\
\hline
E$_0^{++}$ & 0.5732 (5) & 0.679 (1) & 0.782 (1) & 0.886 (2) & 0.987 (2) \\
E$_0^{+-}$ & 1.153 (8) & 1.20 (1) & 1.247 (11) & 1.323 (16) & 1.385 (15) \\
E$_0^{--}$ & 1.547 (8) & 1.555 (8) & 1.581(9) & 1.616 (10) & 1.66 (1) \\
E$_0^{-+}$ & 1.96 (2) & 1.87 (1) & 1.912 (17) & 1.935 (15) & 1.998 (17) \\
\hline
\end{tabular}
\caption{\label{naive} Naive Energies $\beta=5$ Lattice $24^3$}
\end{center}
\end{table}

\begin{table}
\begin{center}
\begin{tabular}{clllll}
\hline
state & $r=4$ & $r=5$ & $r=6$ & $r=7$ & $r=8$ \\
\hline 
E$_0^{++}$ & 0.57119 (5) & 0.6751 (1) & 0.7768 (1) & 0.8777 (1)& 0.9778 (1)\\
E$_0^{*++}$ & 0.5716 (1) & 0.6757 (3) & 0.7776 (5) & 0.8784 (10) & 0.9784 (16) \\
E$_0^{+-}$ & 1.1073 (18) & 1.1429 (25)& 1.1891 (20) & 1.2492 (23) & 1.3165 (22)\\
E$_0^{--}$ & 1.495 (7) & 1.511 (6)& 1.537 (7)& 1.575 (7) & 1.621 (7)\\
\hline
\end{tabular}
\caption{\label{extra24} Extrapolated energies. $\beta=5$. Ground states in different
channels}
\end{center}
\end{table}

\begin{table}
\begin{center}
\begin{tabular}{clllll}
\hline
state & $r=4$ & $r=5$ & $r=6$ & $r=7$ & $r=8$ \\
\hline 
E$_0^{++}$ & 0.5719 (1)& 0.6761 (1) & 0.7782 (2) & 0.8792 (1)& 0.9803 (3)\\
E$_1^{++}$ & 1.626 (31)& 1.665 (20) & 1.704 (33) & 1.766 (6) & \hspace*{5mm} $-$ \\
\hline
\end{tabular}
\caption{\label{samech} Fitted Energies. $\beta=5$. States in \{++\}.}
\end{center}
\end{table}

\begin{table}
\begin{center}
\begin{tabular}{clllll}
\hline
state & $r=6$ & $r=7$ & $r=8$ & $r=9$ & $r=10$ \\
\hline 
E$_0^{++}$ &0.391 (3) &0.436 (5) &0.480 (6) &0.525 (7) &0.568 (8) \\
E$_0^{+-}$ &0.871 (18)&0.903 (23)&0.918 (22)&0.963 (24)&0.984 (21) \\
E$_0^{--}$ &1.171 (6) &1.159 (6) &1.161 (9) &1.167 (9) &1.181 (10) \\
E$_0^{-+}$ &1.53 (2) & 1.49 (2) & 1.52 (2) & 1.52 (2) & 1.55 (2) \\
\hline
\end{tabular}
\caption{\label{naive36} Naive energies. $\beta=7.5$. Ground states in different
channels}
\end{center}
\end{table}

\begin{table}
\begin{center}
\begin{tabular}{clllll}
\hline
state & $r=6$ & $r=7$ & $r=8$ & $r=9$ & $r=10$ \\
\hline
E$_0^{++}$ &0.3825 (1) & 0.4238 (2) & 0.4642 (3)  & 0.5041 (3) & 0.5435 (3) \\
E$_0^{+-}$ &0.716 (9) & 0.729 (7) & 0.735 (7) & 0.741 (11) & \hspace*{5mm}$-$ \\
\hline
\end{tabular}
\caption{\label{extra36} Extrapolated energies. $\beta=7.5$. Ground states 
in different channels}
\end{center}
\end{table}

\begin{table}
\begin{center}
\begin{tabular}{clllll}
\hline
state & $r=8$ & $r=9$ & $r=10$ & $r=11$ & $r=12$ \\
\hline
E$_0^{++}$ &0.308 (3) &0.335 (4) &0.360 (5) &0.387 (5) &0.412 (7)  \\
E$_0^{+-}$ &0.747 (28)&0.764 (31)&0.771 (29)&0.796 (32)&0.808 (31) \\
E$_0^{--}$ &0.986 (20)&0.976 (19)&0.974 (20)&0.973 (20)&0.980 (20) \\
E$_0^{-+}$ &1.354 (13)&1.331 (13)&1.344 (18)&1.344 (14)&1.369 (9) \\
\hline
\end{tabular}
\caption{\label{naive48} Naive energies. $\beta=10$. Ground states in different
channels using T={2,4,6,8}}
\end{center}
\end{table}

\begin{table}
\begin{center}
\begin{tabular}{clllll}
\hline
state & $r=12$ & $r=13$ & $r=14$ & $r=15$ & $r=16$ \\
\hline
E$_0^{++}$ &0.3915 (6) &0.4154 (8) &0.4388 (9) &0.4629 (12) &0.4867 (16)  \\
E$_0^{+-}$ &0.663 (12)&0.681 (15)&0.694 (19)&0.719 (24)&0.733 (30) \\
\hline
\end{tabular}
\caption{\label{naive48a} Naive energies. $\beta=10$. Ground states in $++$ and $+-$
channels using T={4,8,12,16}}
\end{center}
\end{table}

\begin{table}
\begin{center}
\begin{tabular}{clllll}
\hline
state & $r=8$ & $r=9$ & $r=10$ & $r=11$ & $r=12$ \\
\hline
E$_0^{++}$ & 0.2961 (3) & 0.31896 (39) & 0.3419 (6) & 0.36492 (77) & 0.3869 (9) \\
E$_0^{+-}$ & 0.577 (7) & 0.584 (11) & 0.591 (12) & 0.600 (16) & 0.607 (19) \\
\hline
\end{tabular}
\caption{\label{extra48} Extrapolated energies. $\beta=10$. Ground states in different
channels}
\end{center}
\end{table}

\begin{table}
\begin{center}
\begin{tabular}{l|l|l|l}
\hline
$r(r_0)$ & $ \beta $ =5 & $ \beta $ =7.5 & $ \beta $ =10 \\
\hline
1.05 r$_0$ & 0.5264 (21)& 0.3173 (77)& 0.265 (11)\\
1.20 r$_0$ & 0.4849 (31)& 0.287 (7)& 0.245 (13)\\
1.35 r$_0$ & 0.4483 (31)& 0.2543 (87)& 0.226 (18) \\
\hline
\end{tabular}
\caption{\label{ediff} Comparison of the energy difference.}
\end{center}
\end{table}

\end{document}